\documentclass[aps,prx,twocolumn,showpacs,psfig,superscriptaddress,longbibliography]{revtex4-2}

\usepackage{times}
\usepackage{graphicx}
\usepackage{float}
\usepackage{latexsym,amsmath,amssymb,bm,euscript,amsfonts}
\usepackage{color}
\usepackage{subfigure}
\usepackage{epstopdf}
\usepackage{bm}
\usepackage[colorlinks=true,linkcolor=blue,citecolor=blue]{hyperref}
\usepackage{soul}
\usepackage[normalem]{ulem}
\usepackage{mathrsfs}
\usepackage{lettrine}
\usepackage{bbding}
\usepackage{xspace}
\usepackage{textcomp}
\usepackage{textcase}
\usepackage{setspace}
\usepackage{chemformula}
\usepackage{siunitx}
\usepackage{physics}
\usepackage{xfrac}

\def\ECA{EuCo$_2$Al$_9$\xspace}

\begin{document}

%\title{RKKY-Dipolar model and Spin Supersolid in \ECA Alloy}
%\title{Spin Supersolid in an RKKY-dipolar Model for Rare Earth Alloy \ECA }
\title{RKKY-dipolar Interactions and 3D Spin Supersolid on Stacked Triangular Lattice}

\author{Ning Xi}
\thanks{These authors contributed equally to this work.}
\affiliation{Institute of Theoretical Physics, Chinese Academy of Sciences, Beijing 100190, China}
\author{Xitong Xu}
\thanks{These authors contributed equally to this work.}
\affiliation{Anhui Key Laboratory of Low-Energy Quantum Materials and Devices, High Magnetic Field Laboratory (CHMFL), Hefei Institutes of Physical Science, Chinese Academy of Sciences, Hefei, 230031, China}
\affiliation{Science Island Branch of Graduate School, University of Science and Technology of China, Hefei, Anhui 230026, China}
\author{Guoliang Wu}
\thanks{These authors contributed equally to this work.}
\affiliation{Institute of Theoretical Physics, Chinese Academy of Sciences, Beijing 100190, China}
\author{Mingfang Shu}
\thanks{These authors contributed equally to this work.}
\affiliation{Anhui Key Laboratory of Low-Energy Quantum Materials and Devices, High Magnetic Field Laboratory (CHMFL), Hefei Institutes of Physical Science, Chinese Academy of Sciences, Hefei, 230031, China}
\affiliation{College of  Sciences, China Jiliang University, Hangzhou 310018, China}
\affiliation{Key Laboratory of Artificial Structures and Quantum Control, School of Physics and Astronomy, Shanghai Jiao Tong University, Shanghai 200240, China}
\affiliation{Collaborative Innovation Center of Advanced Microstructures, Nanjing University, Nanjing 210093, Jiangsu, China}
\author{Hao Chen}
\affiliation{Lanzhou Center for Theoretical Physics and Key Laboratory for Quantum Theory and Applications of the MoE, School of Physical Science and Technology, Lanzhou University, Lanzhou 730000, Gansu, China}
\author{Yuan Gao}
\affiliation{Institute of Theoretical Physics, Chinese Academy of Sciences, Beijing 100190, China}
\author{Zhentao Wang}
\affiliation{Center for Correlated Matter and School of Physics, Zhejiang University, Hangzhou 310058, China}
\author{Gang Su}
\affiliation{Institute of Theoretical Physics, Chinese Academy of Sciences, Beijing 100190, China}
\author{Jie Ma}
\email{Corresponding authors: jma3@sjtu.edu.cn; zhequ@hmfl.ac.cn; cx@lzu.edu.cn; w.li@itp.ac.cn}
\affiliation{Key Laboratory of Artificial Structures and Quantum Control, School of Physics and Astronomy, Shanghai Jiao Tong University, Shanghai 200240, China}
\affiliation{Collaborative Innovation Center of Advanced Microstructures, Nanjing University, Nanjing 210093, Jiangsu, China}
\author{Zhe Qu}
\email{Corresponding authors: jma3@sjtu.edu.cn; zhequ@hmfl.ac.cn; cx@lzu.edu.cn; w.li@itp.ac.cn}
\affiliation{Anhui Key Laboratory of Low-Energy Quantum Materials and Devices, High Magnetic Field Laboratory (CHMFL), Hefei Institutes of Physical Science, Chinese Academy of Sciences, Hefei, 230031, China}
\affiliation{Science Island Branch of Graduate School, University of Science and Technology of China, Hefei, Anhui 230026, China}
\author{Xi Chen}
\email{Corresponding authors: jma3@sjtu.edu.cn; zhequ@hmfl.ac.cn; cx@lzu.edu.cn; w.li@itp.ac.cn}
\affiliation{Lanzhou Center for Theoretical Physics and Key Laboratory for Quantum Theory and Applications of the MoE, School of Physical Science and Technology, Lanzhou University, Lanzhou 730000, Gansu, China}
\author{Wei Li}
\email{Corresponding authors: jma3@sjtu.edu.cn; zhequ@hmfl.ac.cn; cx@lzu.edu.cn; w.li@itp.ac.cn}
\affiliation{Institute of Theoretical Physics, Chinese Academy of Sciences, Beijing 100190, China}

\date{\today}

\begin{abstract}
Inspired by the recent discovery of metallic spin supersolidity and its giant magnetocaloric effect in the rare-earth alloy \ECA~\cite{Shu2026Nature}, we perform a combined study through electronic structure analysis, effective spin model, and Monte Carlo simulations on a stacked triangular lattice, and reveal a novel mechanism for the emergence of 3D spin supersolid in a metallic antiferromagnet. 
From first-principles inputs, we derive a minimal spin model on a stacked triangular lattice (STL), which arises from the interplay between Ruderman–Kittel–Kasuya–Yosida (RKKY) and dipolar interactions and accurately reproduces the experimental thermodynamics. Based on the STL model, we identify a ground state that simultaneously breaks discrete lattice translational symmetry and continuous spin-rotational symmetry --- the hallmark of a spin supersolid.
Furthermore, we present the field–temperature phase diagram of the 3D STL model and discuss the various magnetic phases and associated phase transitions. 
Under zero field, the spin supersolid Y order establishes in two steps: an upper transition at $T_{N1}$, where an emergent U(1) symmetry appears and the system enters a fluctuating collinear regime, followed by a lower transition at $T_{N2}$ into the spin supersolid Y phase. In contrast, the supersolid V phase undergoes a single phase transition at $T_N^V$. Our results not only provide a comprehensive theoretical understanding of the metallic spin supersolid reported for \ECA but also pave the way for further experimental investigations into its supersolid transitions and universality class.
\end{abstract}
\maketitle

\noindent
\textbf{Introduction.}
The supersolid constitutes an exotic quantum phase of matter that uniquely combines crystalline order with global phase coherence~\cite{Lifshitz1969, Leggett1970, Chester1970, Onsager1956}. The supersolid was pursued experimentally in solid helium~\cite{Leggett1970, Kim2004}, but its existence remained unconfirmed~\cite{Kim2012}. Efforts have since continued to realize such states in ultracold atomic gases~\cite{Li2017Nature, Leonard2017Nature, Tanzi2019, Norcia2021}, and theoretical proposals for its quantum simulation have been put forward in hard-core boson systems on the triangular lattice~\cite{Wang2009, Prokofev2005, Wessel2005, Heidarian2005, Melko2005, Chen2014}. However, the realization of a supersolid phase in condensed matter systems has remained an open question. The spin supersolid~\cite{Sengupta2007field, Sengupta2007chain, Wang2023_SS, Xiang2024Nature, Shu2026Nature, LI2025} constitutes a quantum magnetic analogue of supersolid, with simultaneous breaking of lattice-translation and continuous spin-rotation symmetries. Very recently, spin supersolid has been revealed in triangular-lattice antiferromagnets, such as Co-based compounds such as Na$_2$BaCo(PO$_4$)$_2$ (NBCP)~\cite{Gao2022QMats, LiusuoNBCP, Xiang2024Nature, Gao2024dynamics, Xu2025, Sheng2025Continuum} and K$_2$Co(SeO$_3$)$_2$ (KCSO)~\cite{Zhu2024_KCSO, chen2024_KCSO}.

%%Figure 1
\begin{figure*}[htbp]
%\begin{center}
\includegraphics[width=0.95\textwidth]{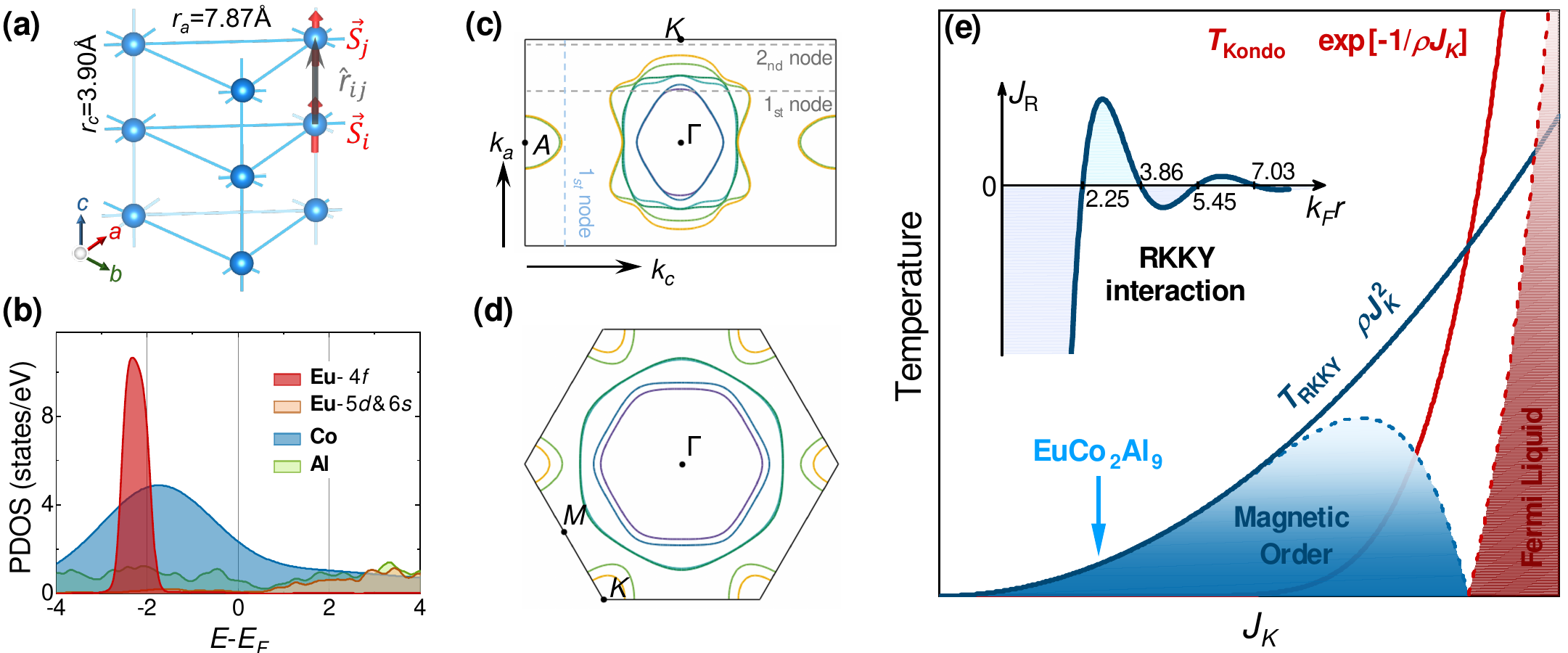}\\[6pt]  % insert figure
\caption{$|~${\bf Electronic structure and localized magnetic moments in \ECA.}
{\bf (a)}, Stacked triangular lattice of Eu$^{2+}$ ions in \ECA (ECA) with interlayer distance ($r_c = 3.901$~\AA) much shorter than the intralayer one ($r_a = 7.871$~\AA). The red arrows refer to the spins on site $i$ and $j$, and the unit vector $\hat{\mathbf{r}}_{ij}$ is also indicated.
{\bf (b)}, Projected density of states (PDOS) of Eu, Co, and Al. 
{\bf (c-d)}, Fermi surfaces of ECA at {\bf (c)} $\Gamma$-$K$-$A$  surface and  {\bf (d)} $\Gamma$-$K$-$M$ surface.  The estimated nodes based on the free-electron approximation along the $k_c$ direction ($\Gamma$–$A$) and $k_a$ direction ($\Gamma$–$K$) are marked in light gray and light blue, respectively. 
{\bf (e)}, Schematic Doniach phase diagram of the Kondo lattice. The blue region where $ T_{\text{RKKY}} > T_{\text{Kondo}} $ indicates the magnetic-ordered regime, and the red region where $ T_{\text{RKKY}} < T_{\text{Kondo}} $ represents the heavy-fermion regime. The boundaries of the two regimes are given by $ |T_{\text{RKKY}} - T_{\text{Kondo}}| $. The ECA is placed deeply at the $ T_{\text{RKKY}} > T_{\text{Kondo}} $ regime. The inset shows the RKKY interaction profile based on the free-electron approximation.
}
\label{fig1}
%\end{center}
\end{figure*}

The spin supersolid exhibits a rich spectrum of excitations~\cite{Gao2024dynamics, Zhu2024_KCSO, chen2024_KCSO, Sheng2025Continuum, Popescu2025, chi2024dynamical, Xu2025, Shi2025, cui2025, huang2025, Chen2026} with intriguing entropic effects~\cite{Xiang2024Nature} and spin transport phenomenon~\cite{Gao2025SSE}. Recent spectroscopic studies have revealed characteristic magnon-roton excitations~\cite{Gao2024dynamics}, closely resembling those in superfluid helium, alongside intermediate- to high-energy excitation continua that may be ascribed to spinon excitations~\cite{Zhu2024_KCSO, chen2024_KCSO, Sheng2025Continuum}. The strong low-energy spin fluctuations give rise to a giant magnetocaloric effect (MCE) at ultralow temperature~\cite{Xiang2024Nature, Shu2026Nature, Xiang2026}, enabling adiabatic demagnetization refrigeration (ADR) down to 100~mK level. Furthermore, the spin supersolid is predicted to host dissipationless spin transport driven by a temperature gradient --- namely, a persistent spin supercurrent in the spin Seebeck effect~\cite{Gao2025SSE} --- underscoring its potential as a versatile platform for exploring macroscopic quantum transport phenomena.

The spin supersolid in the aforementioned Co-based materials, including NBCP~\cite{Zhong2020, Xiang2024Nature, Sheng2025Continuum} and KCSO~\cite{Zhu2024_KCSO, chen2024_KCSO}, emerges in a two-dimensional (2D) triangular lattice easy-axis XXZ model. The effective spin-1/2 arises from strong spin-orbit coupling~\cite{Gao2022QMats, Zhong2020, Ferreira-Carvalho_NBCP}, while the magnetic anisotropy may originate from the octahedral crystal field environment~\cite{Zhong2020, Popescu2025, Wellm2021}. Recently, a spin supersolid was also reported in the Eu$^{2+}$-based alloy \ECA (ECA)~\cite{Shu2026Nature, Xiang2026, Xu2026}. Featuring a stacked triangular lattice, it represents a distinct 3D system for hosting spin supersolidity. 
Unlike Co‑based Mott insulators, ECA is a good metal comprising both localized moments and conduction electrons. The localized moments arise from inner-shell 4$f$ electrons, resulting in weak exchange interactions but a substantial Ruderman–Kittel–Kasuya–Yosida (RKKY) interaction mediated by the conduction electrons. Moreover, owing to its zero orbital angular momentum $L=0$, there is no appreciable spin–orbit anisotropy in ECA. However, due to the large local moment (approximately 7~$\mu_B$) and the small interlayer spacing, interlayer dipolar couplings can give rise to an easy-axis (out-of-plane) magnetic anisotropy. This combination of metallicity, high spin value, and stacked triangular lattice underpins the RKKY–dipolar mechanism for the metallic spin supersolid in ECA~\cite{Shu2026Nature}, distinguishing it from that of previously reported triangular-lattice XXZ antiferromagnets~\cite{Gao2022QMats, Xiang2024Nature}.

In this work, we perform a systematic study of a minimal model for ECA, which we dub the stacked triangular lattice (STL) XXZ model. The model features an intralayer isotropic Heisenberg-type term arising from the RKKY interactions, and an interlayer easy-axis XXZ term originating from both RKKY and dipolar couplings, capturing the 3D nature of the system. Model parameters are determined by fitting to experimental magnetization data, and effectiveness of the model is validated through comparisons and benchmarks against the full interaction model upon the inclusion of longer-range dipolar terms. Based on the minimal STL model, we construct the field–temperature phase diagram, which reveals two distinct spin supersolid phases --- the supersolid Y and V phases. Analysis of the spin supersolid transitions and their universal scaling behaviors reveals distinct critical properties for the Y and V phases. The Y phase exhibits two 3D XY transitions: an upper transition at $T_{N1}$ characterized by an emergent U(1) symmetry, followed by a lower transition at $T_{N2}$ into the Y phase proper. In contrast, the V phase undergoes a single transition at $T_N^V$, where its two order parameters display distinct scaling behaviors. These results open intriguing avenues for future theoretical and experimental exploration.

%~\cite{Zhitomirsky2003, Liu2022QSL}. 
\bigskip
\noindent
\textbf{RKKY interactions in ECA.}
Generally, ECA with both local moments and conduction electrons can be described by the Kondo lattice model~\cite{Fazekas1999,Coleman2015}, which captures the interaction between itinerant electrons and localized spins. As noted by Mott~\cite{Mott1974} and Doniach~\cite{DONIACH1977}, this model involves two competing energy scales: $T_{\text{RKKY}}$ and $T_{\text{Kondo}}$. The former is determined by the RKKY interaction,
\begin{equation}
J_{R}(r) \propto J_K^2\rho\frac{2k_F r \cdot \cos(2k_F r) - \sin(2k_F r)}{(2k_F r)^4},
\label{RKKY}
\end{equation}
provides an effective exchange coupling between local moments mediated by conduction electrons, where $\rho$ is the conduction electron density of states per spin and $J_K$ is the Kondo coupling. By contrast, the Kondo temperature $T_{\text{Kondo}} \sim \exp(-1/\rho J_K)$ ~\cite{Coleman2015} marks the scale below which Kondo screening becomes significant. The competition between RKKY-driven magnetic order and Kondo screening gives rise to the well-known Doniach phase diagram~\cite{DONIACH1977, Fazekas1999,Coleman2015}, illustrated in Fig.~\ref{fig1}{\bf(e)}. When $J_K$ is small, $T_{\text{RKKY}}\sim J_{R}\sim J_K^2$ far exceeds exponentially the suppressed $T_{\text{Kondo}}$, rendering the Kondo scale negligible compared to the RKKY scale between local moments.

Fig.~\ref{fig1}{\bf(b)} presents the DFT-calculated projected density of states (PDOS) of Eu, Co, and Al in ECA, obtained using the Vienna Ab initio Simulation Package (VASP)\cite{KRESSE199615} with the Perdew-Burke-Ernzerhof (PBE) exchange-correlation functional and the PBE+U method (with Hubbard U parameters of 7.1~eV for Eu and 4.0~eV for Co) . The PDOS reveals that itinerant electrons near the Fermi level originate mainly from Co and Al orbitals. In contrast, the Eu $4f$ electrons---which carry the localized magnetic moments---reside in an energy window of about $-3$~eV to $-1.7$~eV, leading to negligible hybridization with the itinerant states.
This weak hybridization is consistent with, and provides a microscopic basis for, the experimental observations.
Experimentally, ECA exhibits clear magnetic phase transitions, yet resistivity measurements show no pronounced Kondo effect down to 300~mK, and a moderate 
specific heat coefficient $\gamma \approx 0.04\ \mathrm{J\cdot mol^{-1}K^{-2}}$ is observed. These results collectively place ECA within the regime where $T_{\text{RKKY}} \gg T_{\text{Kondo}}$, as indicated in Fig.~\ref{fig1}{\bf(e)}. Therefore, in our subsequent modeling, we treat the RKKY interaction as the dominant coupling between local moments, neglecting the Kondo screening effect, i.e., \\
\begin{equation}
\mathcal{H}_R = J_R(r_{ij}) \mathbf{S}_i \cdot \mathbf{S}_j.
\label{heisenberg}
\end{equation}
This RKKY interaction arises from second-order perturbation of the Kondo lattice model. For ECA, we have $J_R/D \sim (J_K/D)^2 \sim 10^{-4}$~\cite{Hayami2017,KOMAROV2017,Ozawa2017} as the RKKY coupling $J_R \sim 0.3$~K and the bandwidth $D \sim 1$~eV. The next-order term is  biquadratic, with $J_{\text{biqu}}/D \sim (J_K/D)^4$, yielding $J_{\text{biqu}}/J_R \sim 10^{-4}$, which is thus safely neglected in our model calculations.

%%Figure 2
\begin{figure*}[t]
\begin{center}
\includegraphics[clip, width=1\textwidth]{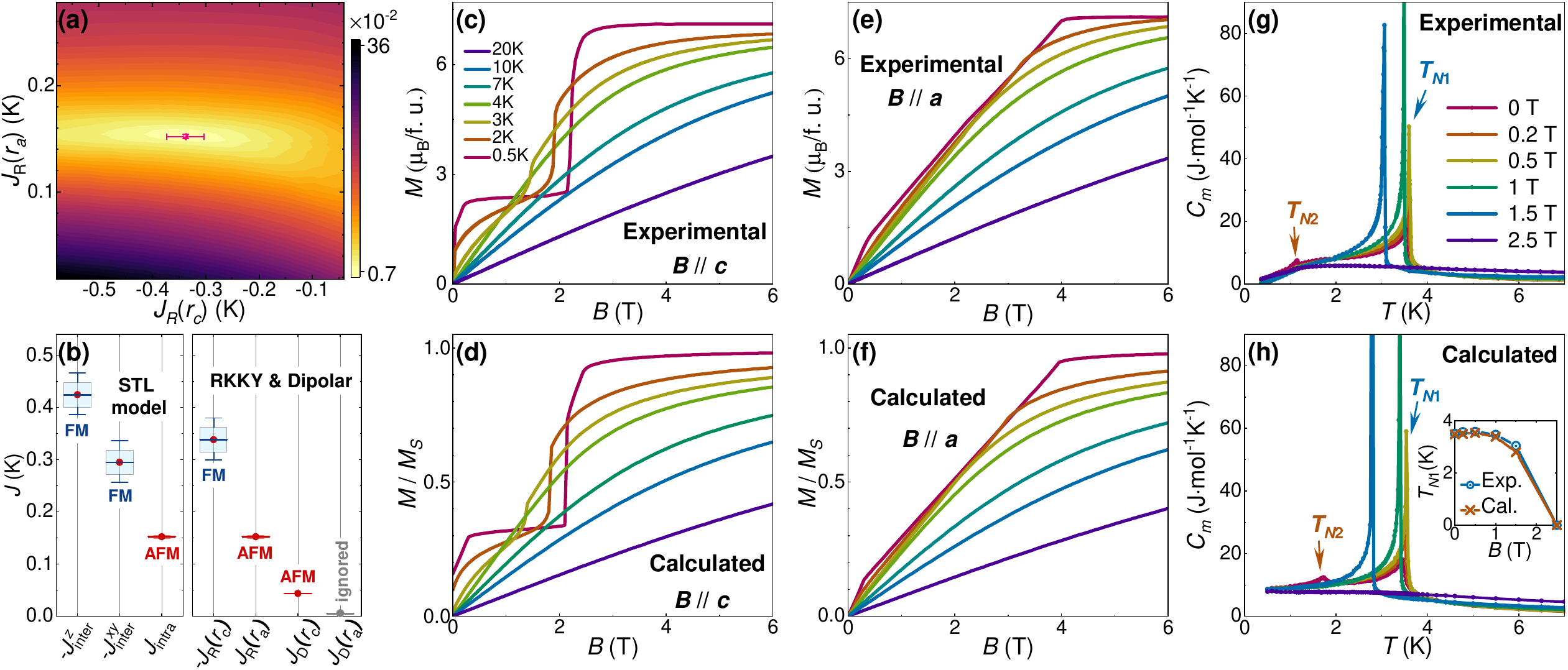} 
\caption{$|~$
{\bf Parameter fitting and comparison between calculated and experimental results. }
{\bf (a)} We fit $J_R(r_a)$ and $J_R(r_c)$ using a loss function involving the out-of-plane magnetization data (see main text), obtaining optimal parameters $ J_R(r_c) = -0.34 \pm 0.04$~K and {$ J_R(r_a) = 0.152 \pm 0.003$}~K.
{\bf (b)} Box plot of optimal parameters. The upper and lower error bars indicate the range of parameters with loss below 0.01, while the top and bottom edges of the box denote the standard deviation of the parameter distribution. The central line within each box represents the mean. Results are shown for the STL model parameters (left) and the RKKY and dipolar interaction parameters (right) for comparison.
{\bf (c-h)} Comparison between calculated {\bf (d, f, h)} and experimental {\bf (c, e, g)}  results: {\bf (c-d)} $c$-axis magnetization,  {\bf (e-f)} $a$-axis magnetization, and  {\bf (g-h)} specific heat for $B\parallel c$. The inset in (h) compares the experimental and calculated upper transition temperature
$T_{N1}$, determined from the prominent specific heat peak, as a function of magnetic field $B$.}
\label{fig2}
\end{center}
\end{figure*}

\bigskip
\noindent
\textbf{Interlayer dipolar couplings and magnetic anisotropy.}
In contrast to Co-based Mott insulators, where the magnetic anisotropy originates from spin-orbit coupling and the octahedral crystal field, the Eu$^{2+}$ ion possesses zero orbital angular momentum ($L=0$) and thus exhibits negligible spin-orbit effects. Its high spin quantum number ($S=7/2$) and large magnetic moment $\sim 7 \mu_B$, however, renders significant magnetic dipole-dipole interactions. Unlike the exchange interaction, the dipolar term can be calculated directly from fundamental crystal parameters, as described by~\cite{Stephen2001,Coey_2010}:
\begin{equation}
\mathcal{E}_D(\mathbf{r}_{ij}) = \frac{\mu_0}{4\pi r_{ij}^3} \left[ \boldsymbol{\mu}_i \cdot \boldsymbol{\mu}_j - 3 (\boldsymbol{\mu}_i \cdot \hat{\mathbf{r}}_{ij}) (\boldsymbol{\mu}_j \cdot \hat{\mathbf{r}}_{ij}) \right],
\end{equation}
where $\boldsymbol{\mu}_i = g\mu_B \mathbf{S}_i$ is the magnetic dipole associated with spin $\mathbf{S}_i$, $\hat{\mathbf{r}}_{ij}$ is the unit vector pointing from site $i$ to $j$ [see Fig.~\ref{fig1}(\textbf{a})], $g$ is the Land\'e factor ($g \simeq 2.03$ for ECA), and $\mu_B$ is the Bohr magneton. This leads to an anisotropic spin-spin interaction of the form:
\begin{equation}
\mathcal{H}_D = J_D(r_{ij})\,
\left[ \mathbf{S}_i \cdot \mathbf{S}_j
- 3 (\mathbf{S}_i \cdot \hat{\mathbf{r}}_{ij})(\mathbf{S}_j \cdot \hat{\mathbf{r}}_{ij}) \right],
\end{equation}
with the dipolar coupling strength $J_D(r_{ij}) = \frac{\mu_0 (g \mu_B)^2}{4\pi r_{ij}^3}.$

\bigskip
\noindent
\textbf{Stacked triangular lattice XXZ model for ECA.}
Based on the foregoing analysis, a microscopic model for the spin structure of ECA must incorporate both the isotropic RKKY interaction and the anisotropic dipolar interaction, i.e.,
\begin{equation}
\mathcal{H} = \mathcal{H}_R + \mathcal{H}_D,
\end{equation}
both long‑ranged and distance‑decaying. The complexity of this spin model necessitates certain simplifications, as detailed below.

The RKKY interaction provides the dominant isotropic exchange. Its oscillatory nature, sketched in the free-electron picture [inset, Fig.~ \ref{fig1}{\bf(e)}], implies that the sign of the coupling depends on the product $k_F r$. 
DFT calculations reveal a multi-sheet Fermi surface [Figs.~\ref{fig1}{\bf (c,d)}]. 
%In Fig.~\ref{fig1}{\bf (c)}, we estimate the nodal positions of the RKKY sign changes within the first Brillouin zone based on the free-electron behavior and the distances $r_a$ and $r_c$. 
In Fig.~\ref{fig1}{\bf (c)},given the free-electron prediction of the RKKY oscillation nodes(e.g., $k_F r = 2.25$ for the first node), we use the interatomic distances $r_a$ and $r_c$ to estimate the nodal positions along $k_F^a$ and $k_F^c$  directions.
This estimation reveals that $k_F^c r_c$ along the $c$-direction lies almost entirely within the first node, whereas $k_F^a r_a$ falls between the first and second nodes. Based on this free-electron estimate, this configuration naturally yields ferromagnetic interlayer coupling along the $c$-axis and antiferromagnetic nearest-neighboring coupling within the triangular layers.
Given the complexity of the actual Fermi surface, in the practical calculation, we treat the two dominant RKKY amplitudes $J_R(r_c)$ and $J_R(r_a)$ as fitting parameters determined by matching Monte Carlo simulations to the experimental magnetization curves.

On the other hand, the dominant anisotropic term arises from the dipolar coupling on nearest-neighbor interlayer bond $r_c = 3.901$~\AA, yielding  $J_D(r_{c}) \simeq 43$~mK and
\begin{equation}
\mathcal{H}_D^{(1)} = J_D(r_c)
\left( \mathbf{S}_i \cdot \mathbf{S}_j-
3 S_i^z S_j^z \right).
\end{equation}
The Eu$^{2+}$ ions form a stacked triangular lattice, as illustrated in Fig.~\ref{fig1}{\bf (a)}, and this term acts as an easy‑axis anisotropy along the crystallographic $c$ direction (spin $z$-axis) while preserving spin rotation symmetry within the $xy$ plane. Longer‑range dipolar contributions are very small: the intralayer nearest distance $r_a = 7.871$~\AA~leads to the dominant symmetry-breaking $J_D(r_{a}) \simeq 5.2$~mK, almost an order of magnitude smaller than $J_D(r_c)$. The magnetic ordering temperature in ECA is approximately 3.5 K, corresponding to an effective exchange scale of about 300 mK. At these energy scales, any weak anisotropic terms --- potentially further washed out by various long-range dipolar couplings --- are negligible in the experimentally relevant sub-kelvin regime and can therefore be safely ignored.

Consequently, by combining the simplified forms of $\mathcal{H}_R$ and $\mathcal{H}_D$, the magnetic behavior of ECA is captured by a STL XXZ model consisting of two distinct parts:
\begin{equation}
\mathcal{H} = \mathcal{H}_{\mathrm{inter}} + \mathcal{H}_{\mathrm{intra}},
\end{equation}
comprising an easy-axis anisotropic interlayer term
\begin{equation}
\mathcal{H}_{\mathrm{inter}} = \sum_{\substack{\langle i,j \rangle_\text{interlayer}}} \left[ J^{xy}_{\mathrm{inter}} (S_i^x S_j^x + S_i^y S_j^y) + J^{z}_{\mathrm{inter}} S_i^z S_j^z \right],
\end{equation}
 and an isotropic intralayer term
\begin{equation}
\mathcal{H}_{\mathrm{intra}} = \sum_{\substack{\langle i,j \rangle_\text{intralayer}}} J_{\mathrm{intra}} \, \mathbf{S}_i \cdot \mathbf{S}_j.
\end{equation}
In this STL model,  the interlayer term $\mathcal{H}_{\mathrm{inter}}$ originates from both the dipolar anisotropy and the RKKY exchange, leading to $J^{z}_{\mathrm{inter}} = -2 J_{D}(r_c) + J_{R}(r_c)$ and $J^{xy}_{\mathrm{inter}} = J_{D}(r_c) + J_{R}(r_c)$. In contrast, $\mathcal{H}_{\mathrm{intra}}$ stems purely from the RKKY interaction, thus $J_{\mathrm{intra}} = J_{R}(r_{a})$. As the two independent parameters are $J_{R}(r_{a})$ and $J_{R}(r_c)$, we fit them and then translate the results into the corresponding STL XXZ model parameters.
\\

\noindent
\textbf{Monte Carlo simulations of the STL model.}
We study the spin models of ECA using classical Monte Carlo simulations, treating spins as classical vectors and employing the Metropolis algorithm for single-spin updates. For the STL model, simulations are performed on lattices of up to $24 \times 24 \times 24$ unit cells. The full STL model includes all long-range dipolar couplings, is treated using the Ewald summation method by separating short- and long‑range contributions in real and reciprocal space on a $6 \times 6 \times 6$ lattice (see Appendix~A). All calculations are carried out with the \textit{SpinToolkit} package~\cite{XuL2025_sptk}. The Monte Carlo simulations reveal a sequential evolution of magnetic configurations along the $B\parallel c$ direction: the Y state, the up-up-down (UUD) state, the V state, and the paramagnetic (PM) state, as shown in Figs. \ref{fig3}{\bf (e-h)}.

In our calculations, we align the spin $z$-axis with the crystalline $c$-axis. For the three-sublattice system, the sublattice magnetizations $\mathbf{m}^{(j)}$ yield the complex order parameters
\begin{equation}
O_\alpha \equiv \frac{1}{\sqrt{3}} \bigl( m^{(1)}_\alpha + m^{(2)}_\alpha \, e^{i 2\pi/3} + m^{(3)}_\alpha \, e^{i 4\pi/3} \bigr), 
\end{equation}
with $\alpha = x, y, z$.
Here, a finite $|O_z|$ reflects the breaking of discrete three-fold (six-fold for zero field) symmetry whereas a finite $|O_{xy}| = \sqrt{|O_x|^2+|O_y|^2}$ signals the breaking of continuous in-plane U(1) symmetry . 
The symmetry breaking can be further revealed by the Monte Carlo sampling distributions of $O_z$ and the in-plane complex vector $m_{xy}=m_x^{(j)} + i m_y^{(j)}$, providing a direct visual signature of the order and its degeneracy.
%%Figure 4
\begin{figure*}[htbp]
\begin{center}
\includegraphics[clip, width=1\textwidth]{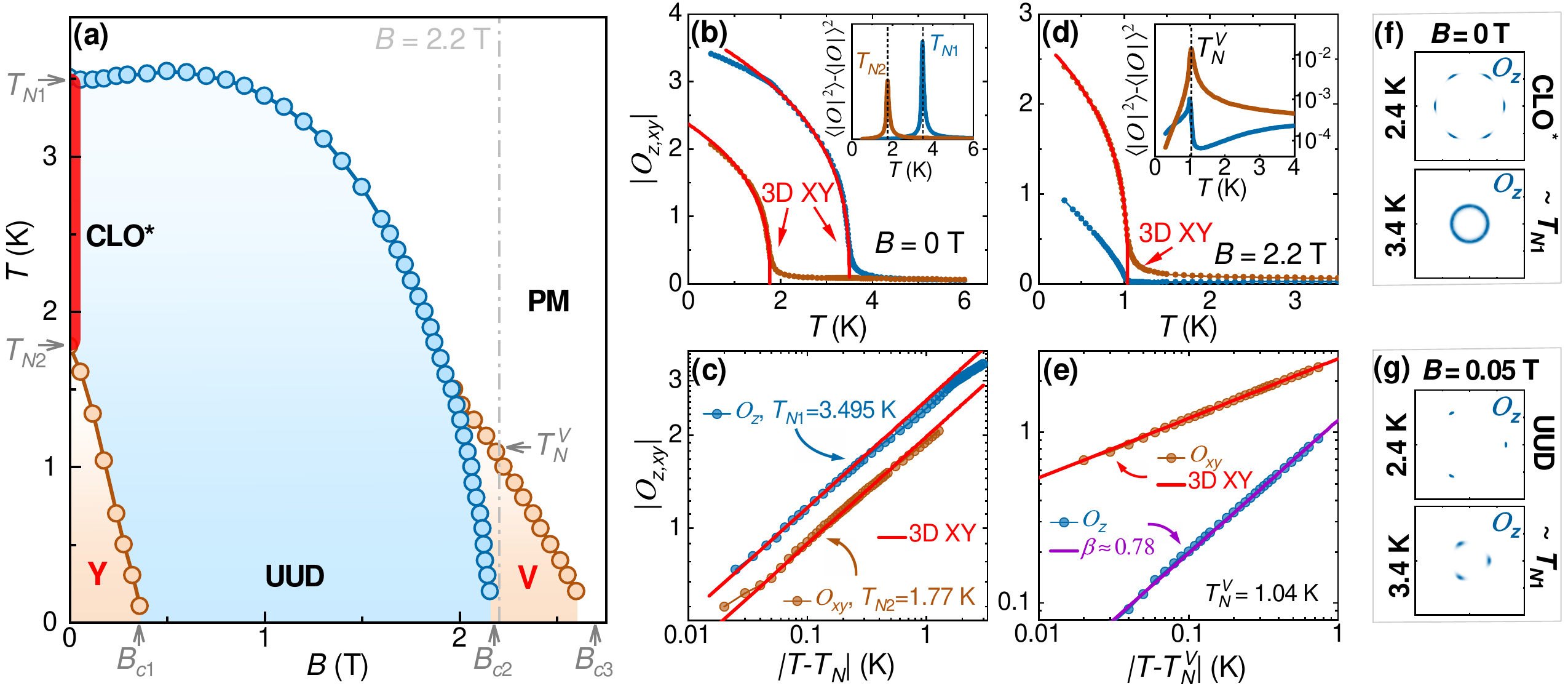}\\[6pt]  % insert figure
\caption{$|~${\bf Calculated phase diagram and order parameters. }
{\bf (a)} The simulated phase diagram of the minimal STL model. The blue and khaki open circles denote the phase boundaries, determined from the peak positions of $\langle |O_z|^2 \rangle - \langle |O_z| \rangle^2$ and $\langle |O_{xy}|^2 \rangle - \langle |O_{xy}| \rangle^2$, respectively. 
{\bf (b,d)} Temperature dependence of the order parameters $|O_z|$ (blue dots) and $|O_{xy}|$ (orange dots) at magnetic fields {\bf (b)} $B = 0$ T and {\bf (d)} $B = 2.2$ T.
{\bf (c,e)} Log-log plot of $|O_z|$ and $|O_{xy}|$ as functions of $T - T_N$ at {\bf (c)} $B = 0$ T and {\bf (e)} $B = 2.2$ T.
The red line in panels {\bf (b–e)} represents the critical behavior of the 3D XY universality class, $(T - T_N)^\beta$, with $\beta \simeq 0.3487$.
In panel {\bf (e)}, the purple solid line corresponds to a power-law fit with $\beta = 0.78$.
{\bf (f, g)} Distribution of the order parameter $O_z$ in the CLO$^*$ phase ($T = 2.4$ K) and near $T_{\text{N}1}$ ($T = 3.4$ K) at {\bf (f)} $B = 0$ T and {\bf (g)} $B = 0.05$ T.
The calculations are performed with parameters $J_R(r_c) = -0.345$ K, $J_R(r_a) = 0.150$ K, and on $24 \times 24 \times 24$ lattice.
%{\bf (e)} Field dependence of the order parameters at a fixed temperature $T = 0.5$ K. A clear jump in the order parameters occurs at the UUD--V phase transition around $2.2$ T. The insets of {\bf (b-e)} show the calculated $\langle |O|^2 \rangle - \langle |O| \rangle^2$.
}
\label{fig4}
\end{center}
\end{figure*}

\bigskip
\noindent
\textbf{Determining model parameters from experimental measurements.}
In the STL model, the dipolar interaction strength $J_D(r_c) \simeq 43$~mK is determined directly from structural parameters and does not require fitting. In contrast, the RKKY couplings $J_R(r_a)$ and $J_R(r_c)$ must be extracted by matching experimental data. We determine these two parameters by fitting the measured $c$-axis magnetization $M_{\mathrm{exp}}$ against the calculated magnetization $M_{\mathrm{calc}}$ from Monte Carlo simulations on an $18 \times 18 \times 18$ lattice at $T = 0.5$, $2$, $3$, and $4$~K.

Fig.~\ref{fig2}{\bf (a)} displays the loss function $\mathcal{L} = \sum_i |M_{\mathrm{exp}}^{(i)} - M_{\mathrm{calc}}^{(i)}| / N$, where $M_{\mathrm{exp}}^{(i)}$ is out-of-plane magnetization data (see Fig.~\ref{fig2}\textbf{c}) and $N$ is the total number of data points. The $\mathcal{L}$ minimum lies near $J_R(r_c) \simeq -0.34$~K and $J_R(r_a) \simeq 0.152$~K. Given systematic uncertainties in the experimental magnetization curves, we regard a loss $\mathcal{L} < 0.01$ as sufficiently accurate estimate. The estimated value of parameters and their bounds of uncertainty are shown in Fig.~\ref{fig2}{\bf (b)}. 

Fixing $J_R(r_c) = -0.345$~K, $J_R(r_a) = 0.150$~K, and $J_D(r_c) = 0.043$~K, we compute the $c$-axis and $a$-axis magnetization curves as well as the field-dependent specific heat from $0.5$~K to $20$~K. The results, compared with experiment in Fig.~\ref{fig2}{\bf (c-h)}, reproduce the measurements remarkably well. We note that the calculated specific heat below $\sim 1.5$~K deviates somewhat from the data, which is expected because our Monte Carlo simulations treat the spins classically and therefore do not capture the full quantum fluctuations at very low temperatures.

\bigskip
\noindent
\textbf{Spin-supersolid phase of ECA.}
In Fig.~\ref{fig4}(a) we present the magnetic phase diagram based on the order parameters $O_z$ and $O_{xy}$ obtained from the minimal model. It shows remarkable agreement with the experimental phase diagram of ECA constructed from various measurements. Schematic real-space spin configurations of the four bulk magnetic phases --- Y, V, UUD, and PM --- are displayed in Fig.~\ref{fig3}{\bf(e-h)} of Appendix.

While Ref.~\cite{Shu2026Nature} inferred the phase diagram from thermodynamic observables, we determine it directly via the order parameters. As shown in Fig.~\ref{fig4}{\bf(b, d)}, both $|O_z|$ and $|O_{xy}|$ are nonzero in the Y and V phases, consistent with a spin-supersolid state that simultaneously breaks translational symmetry and continuous spin-rotation symmetry. In contrast, the UUD phase exhibits finite $|O_z|$ but vanishing $|O_{xy}|$, signaling only discrete symmetry breaking and thus a conventional ``solid-like'' magnetic order.

Besides these four bulk phases, a collinear order (CLO$^*$) phase appears between the zero-field Y and PM phases.  In two-dimensional triangular-lattice systems --- such as NBCP (Refs.~\cite{Xiang2024Nature, Gao2022QMats}) and TmMgGaO$_4$ (Refs.~\cite{Li2020, Hu2020}) --- the CLO$^*$ regime forms an algebraic solid with emergent U(1) symmetry, described as a Berezinskii-Kosterlitz-Thouless phase. In the three-dimensional stacked-triangular lattice of ECA, however, the CLO$^*$ phase breaks the six-fold symmetry, resulting in an ordered state, as shown in the upper panels of Fig.~\ref{fig4}{\bf(f)}: at zero field, the order parameter $O_z$ shows a clear sixfold pattern. 
%This result identifies the CLO$^*$ phase as either the UUD or DDU state, consistent with inelastic neutron scattering observations. 
However, while the experimental CLO$^*$ phase in ECA exhibits zero net magnetization, our calculations of the spontaneously symmetry-broken UUD or DDU states yield a finite net moment---a clear discrepancy. This vanishing magnetization may arise from screening of the local moments by itinerant electrons, an effect beyond the scope of the present STL spin model.

Such sixfold symmetry is predicted to give rise to an emergent continuous U(1) symmetry at the upper critical point\cite{Blankschtein1984}. This is corroborated by the distribution of $O_z$ at $T = 3.4$~K (close to $T_{N1}$) in the lower panels of Figs.~\ref{fig4}{\bf(f)}, which no longer shows six discrete points but rather a uniform, continuous ring in the complex plane --- hallmark of an emergent U(1) symmetry. Furthermore, the critical behavior of $|O_z|$ near $T_{N1}$ follows the 3D XY universality class with exponent $\beta \simeq 0.3487$, as shown in Fig.~\ref{fig4}{\bf(c)}, providing strong evidence for the emergent U(1) symmetry at $T_{N1}$.

Upon further cooling within the zero-field CLO$^*$ phase, the system undergoes a second transition into the Y phase at $T_{N2}$, marked by the onset of the in-plane order parameter $|O_{xy}|$. As shown in Fig.~\ref{fig4}{\bf(c)}, $|O_{xy}|$ acquires a finite value below $T_{N2}$, and its critical behavior near the transition likewise follows the 3D XY universality class, consistent with theoretical expectations.
That is, ECA enters the 3D supersolid Y state via two Bose-Einstein condensations (BEC), analogous to the case in the 2D supersolid compound NBCP~\cite{Gao2022QMats, Xiang2024Nature}, reflecting a common feature of supersolid formation.

In contrast to the two-step transitions into the zero-field Y phase, entering the V phase from the high-temperature disordered state requires only a single transition. As shown in Fig.~\ref{fig4}{\bf(d-e)}, both $|O_{xy}|$ and $|O_z|$ acquire finite values simultaneously at $T_{N}^V$, marking the onset of the V phase. However, their critical behaviors at $T_{N}^V$ differ: $|O_{xy}|$ exhibits the expected critical scaling of the 3D XY universality class near $T_{N}^V$, whereas $|O_z|$ displays a power-law decay with $\beta\simeq 0.78$. Although this indicates a continuous transition, the exponent clearly deviates from the 3D XY value, suggesting that $|O_z|$ at $T_{N}^V$ belongs to a different universality class. 
Based on our current numerical results, the critical behaviors at the supersolid transition $T_{N}^V$ appear to deviate from all known universality classes, and further investigation is required.
To further characterize the supersolid phase transition, we also computed the field-dependent behavior of the order parameters in Appendix~B, which clearly reveals the first-order nature of the transition between the UUD and V states.

%%Figure 3
\begin{figure*}[hbtp]
\begin{center}
\includegraphics[clip, width=0.95\textwidth]{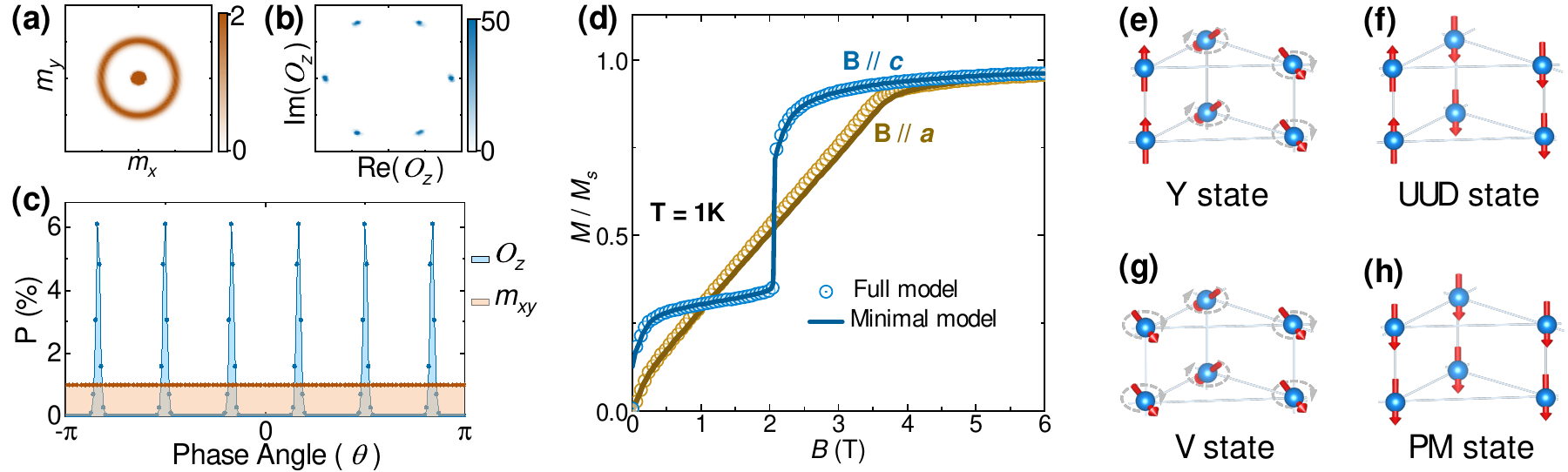}\\[6pt]  % insert figure
\caption{$|~$ {\bf Calculated results of the full STL model with all long-range dipolar interaction included.}
{\bf (a-b)} Calculated distribution of the {\bf (a)} in-plane magnetization $m_{xy} $  and {\bf (b)} out-of-plane magnetic order parameter $O_z$ for the Y state within the full model at $B=0$~T and $T=0.5$~K.
{\bf (c)} Corresponding distribution of the phase angles $\theta$ of $O_z$ and $m_{xy}$. 
{\bf (d)} Calculated magnetization curves for the minimal model and the full model at $T=1$~K .
{\bf (e-h)} Schematic diagrams of four magnetic configurations observed in ECA with a magnetic field $B$ applied along the $c$-axis: {\bf (e)} Y state, {\bf (f)} UUD state, {\bf (g)} V state, and {\bf (h)} paramagnetic (PM) state. 
}
\label{fig3}
\end{center}
\end{figure*}

\bigskip
\noindent
\textbf{Discussion}\\
We have established a theoretical framework for understanding the spin-supersolid phases in ECA through a comprehensive analysis of spin interactions, construction of a minimal model, determination of optimal parameters via Monte-Carlo fitting, and the calculation of order parameters and the corresponding phase diagram. This work reveals a mechanism for 3D metallic spin supersolidity distinct from that in Co-based 2D triangular-lattice systems: a RKKY-dipolar driven spin supersolid. Our numerical simulations based on the minimal STL model --- a combination of RKKY and dipolar couplings --- show excellent agreement with experiments and characterize these magnetic states through order parameters and their scaling behaviors near the supersolid transitions. For the Y phase under zero field, we find a two-step establishment of the supersolid order ($T_{N1}$ and $T_{N2}$); while for the V phase there exist only one phase transition ($T_N^V$). 
Intriguingly, the emergent U(1) symmetry observed at the upper transition $T_{N1}$ places it within the 3D XY universality class --- the same as that of a 3D BEC. As the lower transition at $T_{N2}$ belongs to the same universality class, the spin supersolid state in \ECA is effectively established via two BEC transitions. Beyond this finding, the universality class of the V-phase supersolid transition ($T_N^V$) remains an open question. The STL model thus serves as a rich platform for exploring these critical scaling behaviors --- opening a compelling avenue for future theoretical and experimental investigations.

Within our STL model description, the supersolid Y and V phases break both discrete translational and U(1) spin-rotational symmetries. Although classical Monte Carlo simulations are generally reliable for capturing finite-temperature thermodynamics in large-spin ($S=7/2$) systems like ECA, they inherently neglect quantum fluctuations. This limitation manifests in three ways: discrepancies in specific heat and entropy at low temperatures, a slight overestimation of the transition temperature $T_{N2}$ into the Y phase compared to experiment, and the zero (experiment) vs. nonzero net magnetization (classical model). These deviations highlight the non-negligible role of quantum fluctuations in this frustrated magnet, especially at low temperature. In this sense, we also call for future studies of the STL model with lower spins to emphasize the quantum fluctuating nature of spin supersolidity. This constitutes an intriguing theoretical problem in its own right and holds experimental relevance for other rare-earth (e.g., Yb or Nd, with effective spin-1/2) metallic quantum magnets with similar STL structures.
%The significant magnetocaloric effect reported in Ref.~\cite{Shu2026Nature, Xiang2026} witness the low-temperature fluctuations. 

Moreover, our model considers only the local moments and does not include the enhanced quantum fluctuations due to itinerant electrons. A very recent experiment reveals intriguing interplay between local moments and conduction electrons, as reflected in electrical transport and quantum oscillations~\cite{Xu2026}. Addressing this will require extending our theoretical framework to a more comprehensive model that incorporates both local moments and conduction electrons, thereby accounting for enhanced quantum fluctuations and the contributions of itinerant electrons.Lastly, inspired by the dissipationless spin currents driven by the spin Seebeck effect predicted for the triangular-lattice spin supersolid~\cite{Gao2025SSE}, we expect that ECA may host such supercurrent spin-transport phenomena, awaiting future experimental and theoretical exploration.
\\

%%Figure 5
\begin{figure}[htbp]
\begin{center}
\includegraphics[clip, width=0.8\columnwidth]{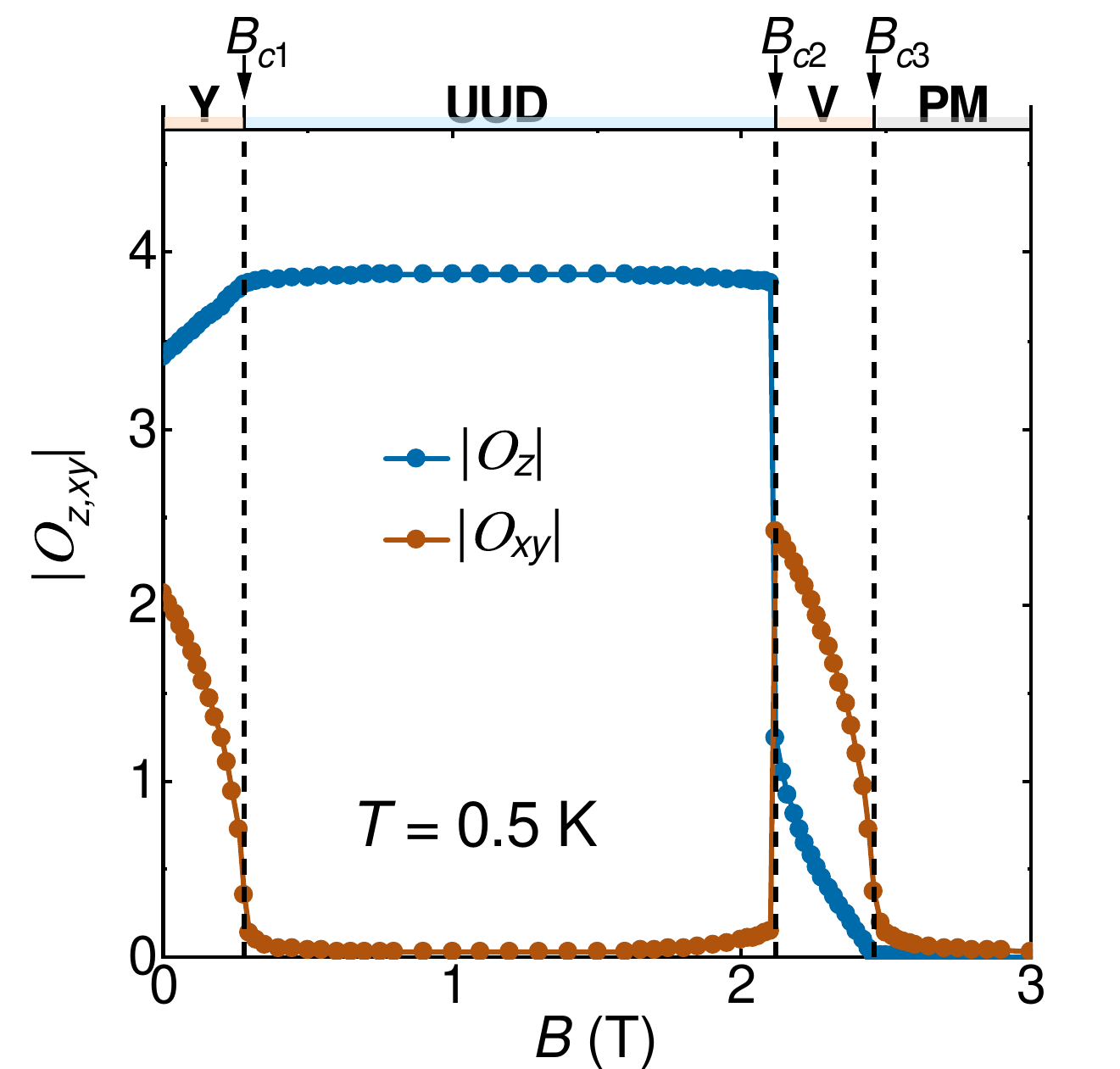}\\[6pt]  % insert figure
\caption{$|~$  Field dependence of the order parameters at a fixed temperature $T = 0.5$ K. The phase transitions from Y to UUD, UUD to V, and V to PM occur at transition fields $B_{c1} \simeq 0.28$~T, $B_{c2} \simeq~2.12$ T, and $B_{c3}~\simeq 2.46$ T, respectively. A clear jump in the both $O_{z}$ and $O_{xy}$ occurs at the UUD-V phase transition around $B_{c2}$.
}
\label{fig5}
\end{center}
\end{figure}

\appendix
\section{Comparison of the Minimal and Full Models}
Our minimal STL model retains only the dominant dipolar term $J_D(r_c)$, which introduces easy‑axis anisotropy along $z$ while preserving the continuous U(1) spin‑rotation symmetry in the $xy$ plane. In the spin supersolid phase, the system break the discrete lattice translation and U(1) spin rotational symmetries. 

On the other hand, although long‑range dipolar couplings exist in principle in \ECA, they are weak and were therefore neglected in the minimal STL model --- except for the nearest-neighbor interlayer coupling. To ensure that the omission of longer-range dipolar couplings is justified and that our conclusions remain robust, we performed Monte Carlo simulations including all dipolar contributions via Ewald summation. It means that we compare (i) the minimal model containing only $J_D(r_c)$ and (ii) the full dipolar model, with fixed RKKY parameters $J_R(r_c) = -0.345$~K and $J_R(r_a) = 0.150$~K.

Fig.~\ref{fig3}{\bf (a)} shows the distribution of the in‑plane sublattice magnetization $m_{xy}$ sampled during the MC run for a $6 \times 6 \times 6$ lattice at $0.5$~K. The distribution exhibits nearly perfect U(1) rotational symmetry in the $m_x$-$m_y$ plane. In contrast, Fig.~\ref{fig3}{\bf(b)} displays the distribution of the out‑of‑plane order parameter $O_z$, which shows six isolated peaks, corresponding to a six‑fold discrete symmetry. The continuous symmetry is further highlighted in Fig.~\ref{fig3}{\bf (c)}, where the phase angle of the complex order parameter $m_{xy}$ is uniformly distributed between $-\pi$ and $\pi$, confirming that the $xy$-plane rotational symmetry remains perfect, even when longer-range dipolar couplings are included.
Moreover, we also compare the magnetization curves obtained from two models. As shown in Fig.~\ref{fig3}{\bf(d)}, the results show quantitative agreement. These findings confirm that the minimal model reliably captures both the spin supersolidity and the associated magnetic phase transitions in ECA.

\section{First-Order UUD--V Phase Transition}
In the main text, we present the calculated phase diagram of the STL model, along with the temperature-dependent behavior of the order parameters for the Y and V phases. As a complementary analysis, we further compute the magnetic-field dependence of the order parameters $O_z$ and $O_{xy}$. As shown in Fig.~\ref{fig5}, the UUD–V transition at low temperature ($T = 0.5$~K) is marked by a discontinuity in the order parameter, indicating a first-order transition. This first-order metamagnetic behavior is further supported by the magnetization jump at the UUD–V transition in Fig.~\ref{fig2}.

\bigskip
\noindent
\textbf{Acknowledgements} \\
This work was supported by the National Key Research and Development Program of China (Grants Nos.~2024YFA1409200, 2024YFA1611101, and 2024YFA1408303), the National Natural Science Foundation of China (Grants Nos.~12504186, 12534009, 12447101, 12374129, 12374124), the Strategic Priority Research Program of Chinese Academy of Sciences (Grant No. XDB1270101), the CAS Project for Young Scientists in Basic Research (Grant No. YSBR-084), and the CAS Project (Grant No. JZHKYPT-2021-08). 
We thank the HPC-ITP for the technical support and generous allocation of CPU time. This work was also supported by Anhui Provincial Major S \& T Project (s202305a12020005), Anhui Provincial Natural Science Foundation No. 2508085ZD013, 2408085J025. A portion of this work was supported by the High Magnetic Field Laboratory of Anhui Province under Contract No. AHHM-FX-2020-02.
\bibliography{ECA-bibliography}
% Produces the bibliography via BibTeX.

\end{document}